\documentclass[prl,twocolumn,showpacs]{revtex4}
\usepackage{bm}
\usepackage{graphicx}

\begin{document}

\title{Amorphous-amorphous transition and the two-step replica symmetry
breaking phase}
\author{Andrea Crisanti$^1$}
\email{andrea.crisanti@roma1.infn.it}

\author{Luca Leuzzi$^{1,2}$} \email{luca.leuzzi@roma1.infn.it}

\affiliation{$1$ Dipartimento di Fisica, Universit\`a di Roma
``La Sapienza'', P.le Aldo Moro 2, I-00185 Roma, Italy}
\affiliation{$2$ Statistical Mechanics and Complexity Center
(SMC) - INFM - CNR, Italy}

\begin{abstract}
 The nature of polyamorphism and amorphous-to-amorphous transition is
investigated by means of an exactly solvable model with quenched
disorder, the spherical s+p multi-spin interaction  model.
The analysis is carried out in the framework of Replica Symmetry
Breaking theory and leads to the identification of low temperature
glass phases of different kinds. Besides the usual `one-step'
solution, known to reproduce all basic properties of structural
glasses, also a physically consistent `two-step' solution arises.
More complicated phases are found as well, as temperature is further
decreased, expressing a complex variety of metastable states
structures for amorphous systems.
 \end{abstract}


\maketitle

In recent years increasing evidence has been collected for the
 existence of amorphous to amorphous transitions (AAT), in various
 glass-forming substances as, e.g., vitreous Germania and Silica,
 where the coordination changes abruptly under pressure shifts
 \cite{Tsiok98}. One refers to this phenomenon as "polyamorphism"
 \cite{Tsiok98,Huang04}.  Like the liquid-glass transition also
 the AAT is not a thermodynamic phase transition, but it amounts to a
 qualitative change in the relaxation dynamics, apparently expressing
 a recombination of the glass structure.  Other kinds of AAT are known
 to occur, e.g.,  in porous silicon \cite{Deb01}, in undercooled water
 \cite{Mishima85,Poole}, in copolymer micellar systems \cite{Chen03}
 and in polycarbonate and polystyrene glassy polymers
 \cite{Cangialosi05}.
\\ \indent A number of theoretical models has been introduced to
describe systems undergoing AAT, as, e.g., a model of hard-core
repulsive colloidal particles subject to a short-range attractive
potential \cite{DawetAl01,Zac01,Sciortino01}.  Another instance is the
spherical $p$-spin model on lattice gas of Ref. \cite{Caiazzo04}.  In
this paper we  consider a model with multibody quenched
disordered interactions where  various AAT's occur as well: the spherical 
$s+p$-spin  model.
\\ \indent Our analytic investigation is performed by applying
Parisi's Replica Symmetry Breaking (RSB) theory \cite{MPV}.  In the
framework of the theoretical description of glasses and, more
generally, of disordered systems, RSB theory provides, in a broad
variety of instances, a rather deep and complex mean-field insight.
RSB solutions so far encountered, representing physically stable
phases, are either one step RSB (mean-field glass) or implement a
continuous hierarchy of breakings (mean-field {\em spin}-glass).  One
step RSB (1RSB) solutions are, e.g., found in the Ising $p$-spin
\cite{Gross84,Gardner85} and Potts \cite{Gross85} models with quenched
disorder in a low temperature interval \footnote{The solution is 1RSB
below the critical temperature at which the paramagnetic phase ceases
to be thermodynamically relevant, but above a second one ("Gardner
temperature") below which it becomes inconsistent.} or in the
spherical $p$-spin model below the static critical temperature
\cite{CriSom92}, else in optimization problems mapped into dilute
spin-glass systems such as the XOR-SAT (where it correctly describes
the whole UNSAT phase) \cite{Montanari03}, or the K-SAT, with $K>2$
\cite{Zecchina99}, in a certain interval of connectivity values next
to the SAT/UNSAT transition \cite{Montanari04}.  The continuous, or
{\em full} (FRSB) solution describes, instead, the low temperature
phase of the mean-field version of the Ising spin-glass, i.e., the
Sherrington-Kirkpatrick model \cite{SheKir75}. Low temperature phases
with a continuous hierarchy of states are known to exist also for
other disordered models with discrete variables, as the
above-mentioned Ising $p$-spin and the Potts model, even though these
might display  a further discontinuous RSB step \footnote{This is
the conjecture of Ref. \onlinecite{Gross85} for the Potts model and
what seems to emerge from preliminary studies of the Ising $p$-spin
model below the Gardner temperature \cite{CriLeuUnp}. As far as we
know a thouroughly computation of the very low temperature phase of
these models has yet to be performed.}.  A purely FRSB phase has been,
eventually, found in a model with continuous variables,
constituted by a two-body and a $p$-body
interaction terms: the spherical $2+p$-spin model \cite{CriLeu04,
CriLeu06}.
\\
\indent
In the present work we consider the spherical $s+p$-spin  model
with both $s$ and $p>2$, addressing physically consistent RSB solutions
qualitatively different from those mentioned above (also different
from the $s=2$ model case).  The
Hamiltonian of the model is
\begin{eqnarray}
\nonumber
&&{\cal H} = \! \sum_{i_1<\ldots <i_s}\!J^{(s)}_{i_1\ldots i_{s}}
\sigma_{i_1}\cdots\sigma_{i_s}
           +\!\sum_{i_1<\ldots <i_p}\!J^{(p)}_{i_1\ldots i_p}
	   \sigma_{i_1}\cdots\sigma_{i_p}
\\
\label{f:Ham}
\end{eqnarray}
where $ J^{(t)}_{i_1\ldots i_{t}}$ ($t=s,p$) are uncorrelated, zero mean, 
Gaussian variables of variance
\begin{equation}
\overline{\left(J^{(t)}_{i_1\ldots i_{t}}\right)^2} = \frac{J_t^2 t!}
{2N^{t-1}}, \qquad
i_1 < \cdots < i_t
\end{equation}
and $\sigma_i$ are $N$ continuous variables obeying the spherical
constraint $\sum_i \sigma_i^2 = N$. Eq. (\ref{f:Ham}) is clearly
symmetric is $s$ and $p$. We will always consider $p>s$. The
properties of the model strongly depend on the value of $s$ and $p$:
for $s=2$, $p=3$ the model reduces to the spherical $p$-spin model in
a field \cite{CriSom92} with a low temperature 1RSB phase, while for
$s=2$, $p\geq 4$ the model posses an additional FRSB low-temperature
phase \cite{Nieuwenhuizen95,CriLeu04} and a 1-FRSB phase
\cite{CriLeu06}.  For $s, p>2$ the phases displayed are the paramagnet
and the 1RSB spin-glass, as far as the difference $p-s$ is not too
large \cite{CriLeu06}. Recently, Krakoviack observed that for large
$p-s$ more than a simple paramagnet/1RSB transition is
likely to occur \cite{Krakoviack}.
\\
\indent
The dynamic equations of the $s+p$ model can be formally rewritten as 
Mode Coupling Theory (MCT) equations in the {\em schematic} approximations, 
see, e.g., Refs. \cite{GoeSjo86,GoeSjo89,KraAlb97,KraAlb02}, and,
defining the auxiliary thermodynamic parameters $\mu_t=t\beta^2J_t^2/2$,
a mapping can be established with the binomial schematic theories 
$F_{s-1,p-1}$ with a scalar kernel, see, e.g., Ref. \cite{CiuCri00}. 
In particular, the $F_{13}$ theory studied 
by G\"otze and Sj\"ogren \cite{GoeSjo89} is dynamically equivalent to a 
$2+4$ spherical spin model \cite{CriLeu07}. 

\section{Large $p-s$ spherical $s+p$-spin model}

Analyzing the model for $s> 2$ and large $p-s$, we observe a very rich
phase diagram even though no purely continuous FRSB phase, as obtained
in the $2+p$ spin model \cite{CriLeu04,CriLeu06}, is encountered in the
present case. We now concentrate on the static scenario.
%
\\
\indent
The static free energy functional reads, for a generic number $R$ of RSB's, 
as \cite{CriLeu06}
\begin{eqnarray}
-\beta \Phi&=& \frac{1}{2}(1+\ln 2\pi) +\lim_{n\to 0}\frac{1}{n}G[{\bm q}] \ ,
\label{f:Phi}
\\
2G[{\bm q}]&=&\sum_{ab}^{1,n}g(q_{ab})+\ln\det {\bm q} \ .
\end{eqnarray}
where 
\begin{equation}
g(q) = \frac{\mu_s}{s} q^s + \frac{\mu_p}{p} q^p
\end{equation}
 and $\bm q=\{q_{ab}\}$ is
the Parisi overlap matrix taking values
$0=q_0<q_1<\ldots<q_R<q_{R+1}=1$. In absence of external
field $q_0=0$.  For any $R$, $G[\bm q]$ can be written as
\begin{eqnarray}
\frac{2}{n}G[{\bm q}] &=& 
            \int_{0}^{1}\!dq\,x(q)\,\Lambda(q) 
            + \int_{0}^{q_R}\!\frac{dq}{\chi(q)}
            + \ln\left(1 - q_R\right)
\nonumber
\\
\quad
\end{eqnarray}
where 
\begin{equation}
x(q)=p_0+\sum_{r=0}^R(p_{r+1}-p_r)\theta (q-q_r)
\end{equation}
 is the
cumulative probability density of the overlaps,
\begin{equation}
\Lambda(q) = \frac{d g(q)}{dq}
\end{equation}
 and
\begin{equation}
\chi(q) = \int_q^1\!dq'~x(q').
\label{f:chi_q}
\end{equation}
Stationarity of $\Phi$ with respect to $q_r$ and $p_r$ leads,
respectively, to the the self-consistency equations
\begin{eqnarray}
{\cal F}(q_r)=0 ,\qquad &\ &r=0,\ldots,R,
\label{f:self_q}
\\
\int_{q_{r-1}}^{q_r}dq~{\cal F}(q) = 0 , 
 &\ & r=1,\ldots,R,
\label{f:self_p}
\end{eqnarray}
where 
\begin{equation}
{\cal F}(q_r)\equiv
\Lambda(q_r)-\int_0^{q_r}\!\frac{dq}{\chi(q)^2}.
\end{equation}
Eq. (\ref{f:self_p}) implies that ${\cal F}(q)$ has at least one
root in each interval $[q_{r-1},q_r]$, that, by the way,
 is not a solution of the
whole set of self-consistency  equations.

\section{How many RSB?}

\begin{figure}[t!]
\includegraphics[width=.49\textwidth]{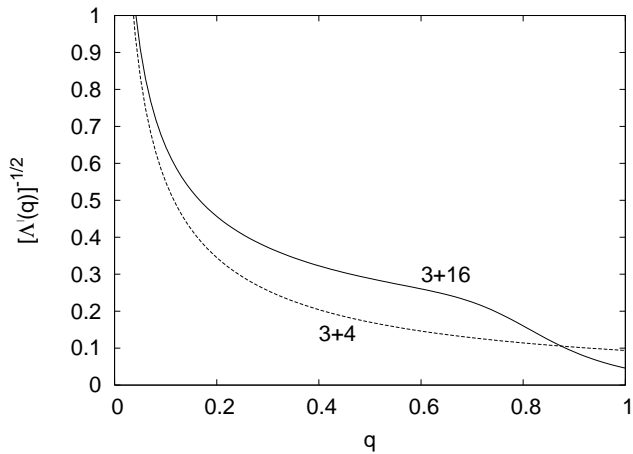}
\caption{Right hand side of Eq. (\ref{f:solutions}) for $\mu_3=12$ and
$\mu_{16}=30$.  The $s+p=3+4$  curve (dotted) can have no more than two
intersections with $\chi(q)$, implying at most a 1RSB solution. This
shape is independent of the particular values of the $\mu$'s (see
Eq. (\ref{f:crit_L}) and discussion thereafter). For $s=3$ and $p=16$
(full curve)
there is, instead, evidence for the existence of a maximum of four
intersections with the concave $\chi(q)$ (because of the double change
of convexity), yielding a 2RSB solution. This does not occur in the
entire plane but only in a sub-region. In the $q$ interval of negative
convexity of $[\Lambda'(q)]^{-1/2}$, $\chi(q)$ can also overlap the
curve in a continuous interval yielding, furthermore, continuous RSB
solutions.}
\label{fig:solutions}
\end{figure}

Which kind of solutions are physically consistent for the 
model at large $p-s$?  
Following Refs. \onlinecite{CriSom92,CriLeu06} we observe that
Eqs. (\ref{f:self_q})-(\ref{f:self_p}) guarantee that between any pair
$[q_{r-1},q_r]$ there must be at least two extremes
of ${\cal F}$ (we recall that here $q_0=0$).
Denoting the extremes by $q^\star$, the condition ${\cal F}'(q^\star)=0$ 
leads to the equation
\begin{equation}
\chi(q^\star)\equiv\int_{q^\star}^1 x(q)~dq=\frac{1}{\sqrt{\Lambda'(q^\star)}},
\label{f:solutions}
\end{equation}
where 
\begin{equation}
\Lambda'(q) = \frac{d \Lambda(q)}{dq}=(s-1)\mu_sq^{s-2}+(p-1)\mu_pq^{p-2}.
\end{equation}  
Since $x(q)$ is a non-decreasing function of $q$, $\chi(q)$ has a
negative convexity. The convexity of the function
$[\Lambda'(q)]^{-1/2}$ depends, instead, on $s$ and $p$, as well as on
the parameters $\mu_s$ and $\mu_{p}$.  If $p-s$ is not too large it
displays a positive convexity in the whole $(\mu_{p},\mu_s)$ plane,
whereas, as $p$ is larger than some critical value that depends on
$s$, the curve can, actually, change convexity in a certain region of
the $(\mu_{p},\mu_s)$ plane.  
\\
\indent
The right hand side of
Eq. (\ref{f:solutions}) is plotted in Fig.  \ref{fig:solutions} in two
qualitatively different model cases with $p>s>2$.  For nearby values
of $s$ and $p$ (dashed curve, $s=3$, $p=4$) the shape of
$[\Lambda'(q)]^{-1/2}$ implies that at most a 1RSB solution can take
place.  When $p-s$ grows, however, the qualitative behavior changes
(full curve, $s=3$, $p=16$) and the 1RSB is no longer the only
solution admissible: solutions with more RSB's may occur in order to
stabilize the system.  From Fig. \ref{fig:solutions} one can readily
see that for certain values of $\mu_s$ and $\mu_p$, and large $p-s$,
Eq. (\ref{f:solutions}) can have four solutions, i.e., ${\cal F}(q)$ can
display four extremes, allowing for the existence of a 2RSB phase.
The critical values of $p$ for fixed $s>2$, above which this kind of
phase can show up, are those at which the $[\Lambda'(q)]^{-1/2}$ function
acquires a negative convexity at some given $q$ value between $0$ and
$1$, i.e., the values for which
\begin{equation}
(p^2 + p + s^2 + s - 3 s p)^2 - p s (p-2) (s-2) = 0
\label{f:crit_L}\end{equation} 
e.g., $(s,p)=(3,8)$, $(4,7+2\sqrt{6})$, $(5,9+3\sqrt{5})$.  
As $s$ increases, the relative critical $p$ becomes very large
\footnote{For the expert reader we mention that the values of $p$
corresponding to a certain $s$ are lower than those provided by
Krakoviack \cite{Krakoviack}. Those, indeed, correspond to the values
of $p_{d}(s)$ for which the dynamic transition line between the
paramagnetic phase and the frozen 1RSB phase develops a
swallowtail. Some examples are $(s,p_d(s))=(3,10)$, $(4,15.92)$,
$(5,21.79)$.  Since the swallowtail is a signal that the 1RSB is
becoming unstable somewhere it is a sufficient condition to have a $p$
larger than or equal to $p_d$ to know that more RSB's are needed. The
appearence of the swallowtail is not, however, a necessary
condition. To complete the picture, we also mention that a swallowtail
develops in the static PM/1RSB transition line as well, at a still larger
$p_s(s)>p_d(s)$. For instance $(s,p_s(s))=(3,12.43)$, $(4,22.68)$,
$(5,34.24)$.}. Notice that Eq. (\ref{f:crit_L}) does not depend on the
parameters $\mu$. For $p(s)$ less than the root of the equation,
$[\Lambda'(q)]^{-1/2}$ has always a positive convexity (see the $3+4$
curve in Fig. \ref{fig:solutions}), while for larger values of $p$ its
convexity can be negative for some renge of values of the $\mu$'s.
\\
\indent
Eqs. (\ref{f:self_q})-(\ref{f:self_p}) also admit continuous RSB
solutions ($q_r-q_{r-1}\to 0$). They reduce to the same identity in
this case, as well as ${\cal F}'(q)=0$ [Eq. (\ref{f:solutions})]. The
latter sets a constraint on the interval of $q$ values over which a
FRSB Ansatz can be constructed because Eq. (\ref{f:solutions}) can
only hold as far as $[\Lambda'(q)]^{-1/2}$ has the same (negative)
convexity as $\chi(q)$. A continuous RSB
structure in a certain interval of $q$ values does not rule out,
however, the possibility of discrete RSB's in other intervals. Indeed,
`mixed' solutions are found as well, whose overlap function $q(x)$, inverse
of $x(q)$,
both display discontinuous steps and a continuous part.

\section{Phase diagram}

\begin{figure}[t!]
\includegraphics[width=.48\textwidth]{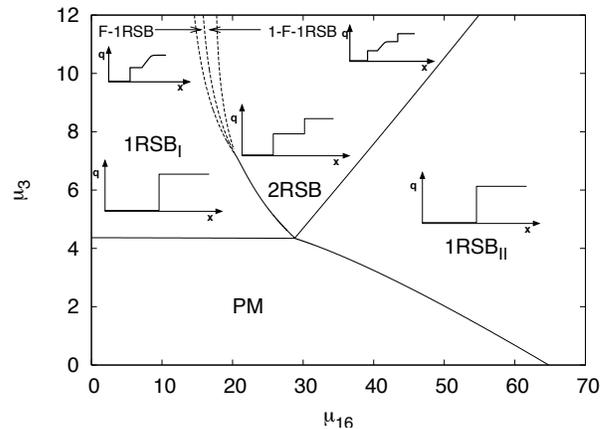}
\caption{Static ($\mu_p$, $\mu_s$) phase diagram of the spherical
$3+16$ spin glass model.  PM: Paramagnetic phase; 1(2)RSB: one(two)
step(s) RSB phase; F-1RSB: full RSB with one discontinuous step at low
$q$: 1-F-1RSB; full RSB terminating both with a step at small and large $q$.
  Overlap order parameter functions are drawn in the
various phases. }
\label{fig:phdi.3+16}
\end{figure}

\begin{figure}[t!]
\includegraphics[width=.47\textwidth]{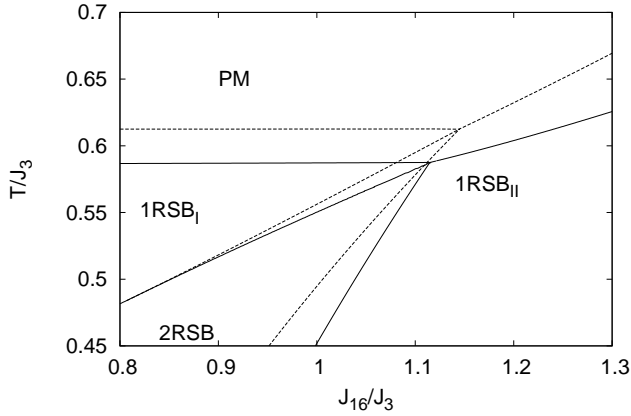}
\caption{Detail of the $T-J_{16}$ phase diagram around the static
(continuous) and dynamic (dashed) quadricritical points. Temperature
and $J_{16}$ are in units of $J_3$. Starting from any phase in the
diagram lowering the temperature the system first undergoes the
dynamic transition and then the statics.  }
\label{fig:phdi.3+16.TJ}
\end{figure}

\begin{figure}[b!]
\includegraphics[width=.49\textwidth]{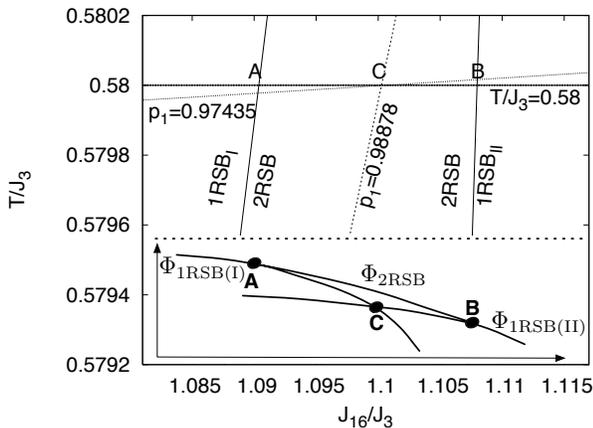}
\caption{Phase transitions at fixed $T=0.58$ varying $J_{16}$ in $J_3$ units.
Point C is the (avoided) first order phase transition, A and B 1RSB/2RSB 
transition points. Bottom: qualitative picture of the behavior of the 
isothermal free energy versus $J_{16}$.
As $J_{16}$ is increased the globally stable thermodynamic phase passes from
a 1RSB to a 2RSB solution and back to a 1RSB one.}
\label{fig:free_cross}
\end{figure}

\begin{figure}[t!]
\includegraphics[width=.49\textwidth]{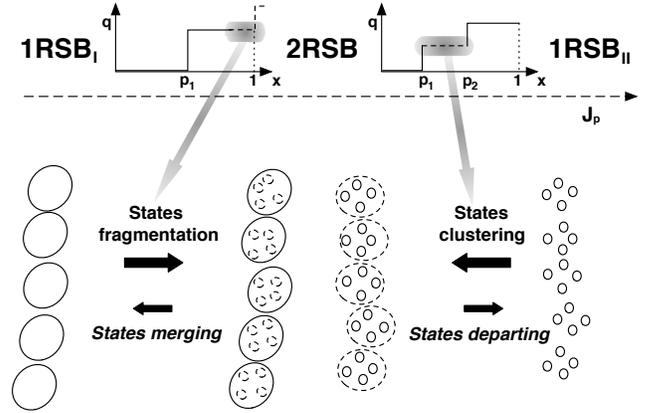}
\caption{Qualitative change of the overlap function across the
1RSB$_I$/2RSB and the 2RSB/1RSB$_{II}$ transitions.
The first transition occurs as a second step at larger $q$ appears at
$p_2=1$.  The second one takes place as the intermediate step between
$p_1$ and $p_2$ shrinks and vanishes.}
\label{fig:q_x}
\end{figure}

We now inspect the explicit case where $s=3$ and $p=16$, that is
enough to catch the properties that make the model special, without
losing of generality.  The complete phase diagram is plotted in the
MCT-like variables $\mu_3$, $\mu_{16}$ in Fig. \ref{fig:phdi.3+16},
where we only report the static transition lines. We stress, however,
that all static transitions have a dynamic counterpart, as we will
discuss in a later section.  In each phase, the shape of the overlap
function $q(x)$ is sketched.  Going clockwise, in the central part we
identify: a paramagnetic (PM) phase, a 1RSB glassy phase ($I$), a 2RSB
glassy phase and a second 1RSB phase ($II$).  Even though the
structure of the states organization is qualitatively similar, the two
1RSB phases differ in the value of the self-overlap $q_1$ (or
Edwards-Anderson parameter \cite{EdwAnd75}) and the position $p_1$ of the RSB
step along the $x$-axis.  In the top part of the phase diagram things
get even more diversified and we find the additional mixed
continuous-discontinuous ``F-1RSB'' (shaped as in the top-left inset
of Fig. \ref{fig:phdi.3+16})\footnote{This kind of order parameter
function was first conjectured for the lowest temperature spin-glass
phase of the mean-field disordered Potts model \cite{Gross85}.}  and
``1-F-1RSB'' phases.
\\ \indent In Fig. \ref{fig:phdi.3+16.TJ} a detail of the phase
diagram is plotted around the quadricritical point where four
transition lines meet.  We use in this case the natural thermodynamic
parameters $T$ and $J_p$ (in units of $J_s$) rather than the MCT
parameters $\mu_s, \mu_p$.  The dynamic transition curves are also
plotted (dashed lines) in this case.  We notice that decreasing the
temperature the dynamic transition always takes place before the
static one.
\\
\indent
Starting in the 1RSB$_{I}$ phase for low values of the ratio
$J_p/J_s$, if we increase $J_p$ keeping temperature fixed, at some
point A, see Fig. \ref{fig:free_cross},
a 2RSB phase arises with the same free energy of the
1RSB$_{I}$. As $J_p$ is further increased,
the 2RSB phase displays a higher free energy than the 1RSB$_I$ one
(bottom inset of Fig. \ref{fig:free_cross}).  Since we are considering
replicated objects in the limit of the number of replicas going to zero, this
implies that the 2RSB phase is the stable one, whereas the 1RSB phase
becomes metastable. 
In Fig. \ref{fig:free_cross} we show the detail of
the  1RSB$_I$-2RSB-1RSB$_{II}$ isothermal transitions in $J_p/J_s$.  In
the inset we show the free energies $\Phi(T/J_s,J_p/J_s)$ relative to
each phase.  It is clear that, would the 2RSB phase not be there, a
coexistence region would occur, as well as a related first order
1RSB$_{I}$/1RSB$_{II}$ phase transition (point C).  The rising of the
2RSB phase covering the whole interested region, however, prevents
 the occurrence of a first order phase transition.
\\ \indent In terms of states reorganization, or order parameter
functional shape, the way the system undergoes the 1RSB$_{I}$/2RSB and
the 2RSB/1RSB$_{II}$ transitions is not the same, as it is reported in
Fig.  \ref{fig:q_x}.  The first transition is a straightforward
generalization of the $p$-spin model PM/1RSB transition (also taking
place in the presenet model, see Figs. \ref{fig:phdi.3+16} and
\ref{fig:phdi.3+16.TJ}): the second step appears, indeed, at $x=1$ as
the highest one.  This means that new states arise {\em inside} the
states of the 1RSB phase, while the latter acquire the status of
clusters of states. We will refer to this kind of transition as {\em
states fragmentation} \cite{RivBir04}.  Across the second kind of
transition, instead, going from the 1RSB$_{II}$ to the 2RSB phase, an
intermediate step of $q(x)$ appears at $p_1$.  This corresponds to
group the states into clusters whose relative overlap is equal to the
value of the intermediate step - between $p_2$ and $p_1$ in the right
hand side overlap picture in Fig.  \ref{fig:q_x} - i.e., the system
undergoes a sort of {\em states clustering} transition. Precursory
symptoms of these transitions were observed in a glassy model on Bethe
lattice in Ref. \onlinecite{RivBir04} where the 1RSB solution of the
model proved unstable in the close packing limit and two kinds of
instability where considered, leading to the onset of a phase
described by a more refined RSB Ansatz.  Here we have an explicit
realization of that conjecture.

\subsection{Formal characterization of phases and transitions}
\begin{figure}[t!]
\includegraphics[width=.4\textwidth]{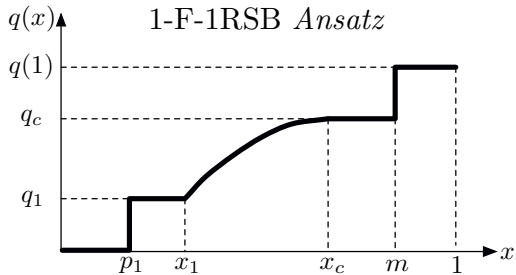}
\caption{Picture of the 1-F-1RSB solution and relative notation for the
overlap values ($q$-axis) and the replica symmetry breaking parameter values 
($x$-axis).}
\label{fig:1f1rsb}
\end{figure}

In order to describe the various phases and their relative
transitions, one has to solve the saddle point
Eqs. (\ref{f:self_q})-(\ref{f:self_p}). In introducing the RSB
solutions describing the frozen phases of the spherical $s+p$ model
for large $p-s$, we start from the most complicated one that we have
found: the 1-F-1RSB phase.  From the equations describing this one, 
indeed, the self-consistency equations of all the other phases can
be straightforward derived, as well as the phase transition lines.
\\ \indent In general, we will denote by $q_1$ the value of the first
step in the overlap function $q(x)$ and by $p_1$ its position on the
$x$ axis. The last step will be identified by $q(1)$, i.e., the value
of $q(x)$ in $x=1$, and the relative step position by $m$. If a
continuous part is present, we  call its highest overlap value
$q_c$, reached at $x_c=x(q_c)$. The initial point on the $x$ axis of
the continuous part is $x_1=x(q_1)$.  In Fig. \ref{fig:1f1rsb} a
pictorial plot of the 1-F-1RSB $q(x)$ is shown, to help fixing the
notation.

The self-consistency equations, cf. Eq. (\ref{f:self_q}), 
are expressed, for this solution, by
\begin{eqnarray}
\Lambda(q_(1))-\Lambda(q_c)&=&\frac{q(1)-q_c}{\chi(q(1))\chi(q_c)}
\label{f:eq_1f1_R}
\\
\Lambda(q)-\Lambda(q_1)&=&
\int_{q_1}^q\frac{dq}{\chi^2(q)}, \qquad q_1\leq q\leq q_c
\label{f:eq_1f1_con}
\\
\Lambda(q_1)&=&
\frac{q_1}{\chi(0)\chi(q_1)}
\label{f:eq_1f1_1}
\end{eqnarray}
where $\chi(q)$ is given in Eq. (\ref{f:chi_q}) and in the above cases 
(no external field, $q_0=0$) it
takes the expressions 
\begin{eqnarray}
\chi(q(1))&=& 1-q(1)
\\
\chi(q_c)&=& 1-q(1)+m[q(1)-q_c]
\\
\chi(q_1)&=& 1-q(1)+m[q(1)-q_c] + \int_{q_1}^{q_c}\!dq~x(q)
\\
\chi(0)&=& 1-q(1)+m[q(1)-q_c] + \int_{q_1}^{q_c}\!dq~x(q) + p_1 q_1
\nonumber
\\
\end{eqnarray}
 Using the function
\begin{equation}
z(y)=-2y\frac{1-y+\log y}{(1-y)^2}
\label{f:z_y}
\end{equation}
introduced by Crisanti and Sommers (CS) in
Ref. \onlinecite{CriSom92}, the self-consistency equation for the RSB points
$p_1$ and $m$,
cf. Eq. (\ref{f:self_p}), become
\begin{eqnarray}
z(y(1))&=&2~\frac{g(q(1))-g(q_c)-[q(1)-q_c]\Lambda(q_c)}{[q(1)-q_c]
[\Lambda(q(1))-\Lambda(q_c)]}
\label{f:eq_1f1_yR}
\\
z(y_1)&=&2~\frac{g(q_1)}{q_1\Lambda(q_1)}
\label{f:eq_1f1_y1}
\end{eqnarray}
with 
\begin{equation}
y_1=\frac{\chi(q_1)}{\chi(0)}, \qquad y(1)=\frac{\chi(q(1))}{\chi(q_c)}.
\end{equation}
\\ \indent From this phase the system can undergo two transitions:
toward a F-1RSB phase, on the left hand side in the
$(\mu_{16},\mu_{3})$ plane, and toward a 2RSB phase, on the right hand
side (see Fig. \ref{fig:phdi.3+16}).  Transforming into the 2RSB
phase, $q_c\to q_1$ and the continuous part disappears.  The saddle
point equations left to yield the solution are Eqs. (\ref{f:eq_1f1_R})
and (\ref{f:eq_1f1_yR}), with $q_c=q_1$, together with
Eqs. (\ref{f:eq_1f1_1}) and (\ref{f:eq_1f1_y1}). We might say that we
are facing a state {\em departing}, i.e., the opposite of the state
clustering mentioned in the previous section and in the r.h.s. of
Fig. \ref{fig:q_x}). This, now, takes place on a continuous set of
ultrametric levels, eliminating the whole intermediate structure of
clusters of clusters, and eventually leaves a three level organization.
On the left hand side, instead, the F-1RSB/1-F-1RSB
transition occurs as $q(1)= q_c$ and Eqs. (\ref{f:eq_1f1_R}) and
(\ref{f:eq_1f1_yR}) become trivial identities. Coming from the F-1RSB
side, states break down into smaller states, themselves becoming
clusters of these newborn states. This is a fragmentation transition, cf.
l.h.s. of Fig. \ref{fig:q_x}) and Ref. \cite{RivBir04}.  
\\ \indent We now continue scanning the phase diagram of
Fig. \ref{fig:phdi.3+16} counterclockwise. Decreasing the variances of
the random coupling distribution at constant temperature, or else
increasing the temperature at fixed interaction variances, the system
in the F-1RSB phase ends up into a 1RSB frozen phase. At the
transition $q_c\to q_1$ and the continuous part is suppressed
\footnote{This transition may occur, as well, if $x_1\to 1$, but this
does not take place in reality in the present model.}. The whole
ultrametric continuous structure inside the largest clusters merges
into simple states.
\\ \indent Lowering $\mu_3$ (i.e. increasing the temperature or
decreasing $J_3$) the system reaches the paramagnetic phase at which
$q_1$ jumps to $0$ discontinuosly, because $p_1$ overcomes $1$, and
only one state remains.  Increasing $\mu_{16}$, the paramagnet goes
back to a frozen - multistates - glassy phase, the 1RSB$_{II}$, as a
step $q_1$ appears at $x=p_1=1$. These last two are transitions of the
kind occurring in the spherical $p$-spin model in which the
Edwards-Anderson order parameter discontinuously jumps from zero to
$q_1$. They are also termed random first order transitions
\cite{KirTirWol89}.
\\ \indent From the 1RSB$_{II}$ phase the system goes into a 2RSB
phase in a states clustering transition, as already mentioned above
and shown in Fig. \ref{fig:q_x} (right hand side).  Eventually, from
the 2RSB the system can transform into the 1RSB$_I$ phase (left hand
side of Fig. \ref{fig:q_x}) or into our starting phase, the 1-F-1RSB
phase.

\section{Dynamic transitions}
\begin{figure}[t!]
\includegraphics[width=.49\textwidth]{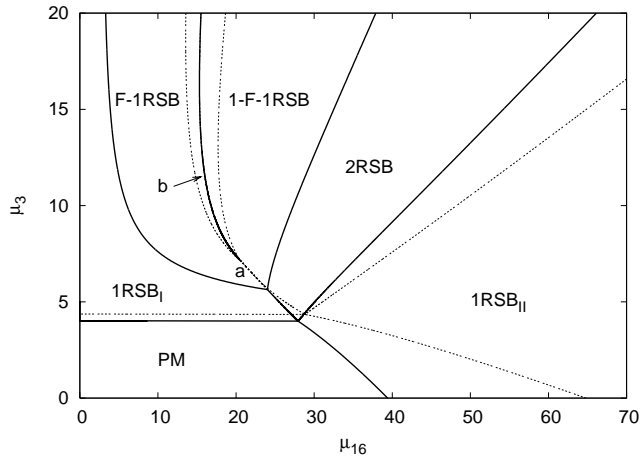}
\caption{Dynamic and static (dotted) transition lines in the
$(\mu_{16}, \mu_3)$ plane. The dynamic lines are all plotted as
continuous curves but for [a] (dashed), which denotes the
discontinuous part of the 1RSB$_I$/F-1RSB transition. By [b] we
indicate the continuous part of this transition where statics and
dynamics coincide.}
\label{fig:dyn_PhDi}
\end{figure}

For what concerns the dynamic glass transitions (dashed lines in
Fig. \ref{fig:phdi.3+16.TJ}), they can be obtained looking at the
dynamical solutions as formulated in Ref. \onlinecite{CriLeu07}, where
the equilibrium dynamics of the system in the different regions of the
parameters space was analyzed and the solution for a generic number of
different relaxation times was provided.  In the dynamic-static
analogy initially proposed by Sompolinsky \cite{Sompolinsky81}, each
of the relaxation time bifurcations corresponds to a RSB.  The
dynamic solution equivalent to the static 1-F-1RSB phase is described
by the Eqs.  (\ref{f:eq_1f1_R})-(\ref{f:eq_1f1_1}) plus the so-called
{\em marginal conditions}
\begin{eqnarray}
\Lambda'(q(1))&=&\frac{1}{\chi^2(q(1))}
\label{f:dyn_1f1_R}
\\
\label{f:dyn_1f1_1}
\Lambda'(q_1)&=&\frac{1}{\chi^2(q_1)} \ .
\end{eqnarray}
The dynamical solution leads, moreover, to the further identity
\begin{equation}
x(q_1)=p_1
\label{f:dyn_p1}
\end{equation}
We notice that the solution is not overdetermined since, 
because of the presence of the continuous part in $q(x)$,
Eq. (\ref{f:dyn_1f1_1}) is not independent from Eq. (\ref{f:eq_1f1_con}).
\\
\indent
The dynamic phase diagram is plotted in Fig. \ref{fig:dyn_PhDi}
where the static lines are also reported (dotted curves) for a direct
comparison with the diagram of Fig. \ref{fig:phdi.3+16}. As one can
see, each static solution representing an equilibrium phase has its
dynamic counterpart. The only difference is that the transition line
between the F-1RSB and the 1-F-1RSB phases can be both continuous (b) and
discontinuous (a), whereas in the static case it was only continuous.
The transition lines are, then,  obtained from the dynamic solutions.
In table \ref{tab:dyn_trans} we report, in the current notation
for the overlap  and the RSB parameter values, the description of all the
phases involved and the relative transitions. 

Alternatively, one can determine the dynamic transitions starting from
the analysis of the behavior of the static functions. The
discriminating quantity is not the free energy, in this case, but the
total complexity of the states \cite{Monasson95,CriSom95}, that is the
average over the quenched disorder of the logarithm of the number of
metastable states.  Indeed, to see which phase is relevant at a given
phase diagram point, one has to select the one with the higher total
complexity and the loci where the complexities of the different phases
equal each other are, thus, the transition lines reported in
Fig. \ref{fig:dyn_PhDi}.
 The complexity can be obtained, e.g.,  as the
Legendre transform of the free energy $\Phi$, cf. Eq. (\ref{f:Phi}) with
respect to the parameter $m$, i.e., the last breaking point,
corresponding to the {\em states level} in the ultrametric tree (see
Fig. \ref{fig:q_x}).
\\ \indent We, eventually, notice that, in Fig. \ref{fig:dyn_PhDi}, in
one case the dynamic line conicides with the static one, at the
F-1RSB/1-F-1RSB transition. This is typical of phase transitions where
the $q(x)$ ends with a continuous part [$q(1)=q_c$] on one side and
with a discontinuous step on the other side of the transition line
(that at $x=m$ in our notation).  If the step is on top of a
continuous part and goes to zero {\em smoothly}, that is, continuously,
at the transition (as in the present case) or if a continuous part
smoothly develops on top of the step
(as it is conjectured to occur in other models, see, e.g.,
Ref. \onlinecite{Gross85}), this implies that the complexity of the
system goes to zero as well
\footnote{We are speaking about the so-called BRST, or
``supersymmetric'', complexity counting states represented by minima
of the free energy landscape, see Refs. \onlinecite{Mueller06} and
\onlinecite{Parisi06} and references therein for a broadening on the
subject.}  and the dynamics does not get stuck at some excited state,
thus reaching the thermodynamic static solution. The same is known to
happen, e.g., in the $2+p$-spin version of the model
\cite{CriLeu06,CriLeu07} at the 1RSB/1-FRSB transition
and in the Ising $p$-spin model
\cite{Crisanti05,CriLeuUnp} at the 1RSB/FRSB transition.

\begin{widetext}
\begin{center}
\begin{table}[t!]
\begin{tabular}{|l|c|c|c|c|c|c|}
\hline
Solution &1-F-1RSB &F-1RSB &\phantom{F}1RSB$_I$\phantom{1F} &\phantom{FF1}PM\phantom{FF1} &\phantom{F}1RSB$_{II}$ \phantom{F}&\phantom{1F}2RSB \phantom{1F}
\\
\hline
1-F-1RSB & &$m=1$ [a]& & & &$q_c=q_1$ 
\\
 $[p_1,q_1,x_c,q_c,m,q(1)]$\, & &\, $q(1)=q_c$ [b]\, & & & &
\\
\hline
F-1RSB  & $m=1$ [a]& &$q(1)=q_1$ & & &
\\
 $[p_1,q_1,m,q(1)]$\, & \, $q(1)=q_c$ [b] \, & & & & &
\\
\hline
1RSB$_I$ & &$q(1)=q_1$ &  &$p_1=1$ & &$p_2=1$
\\
$[p_1,q_1]$ & & & & & &
\\
\hline
PM & & &$p_1=1$ & &$p_1=1$ &
\\
$[q(x)=0~ \forall x]$ & & & & & &
\\
\hline
1RSB$_{II}$ & & & &$p_1=1$ & &$p_2=p_1$
\\
$[p_1,q_1]$ & & & & & &
\\
\hline
2RSB &$q_c=q_1$ & &$p_2=1$ & &$p_2=p_1$ &
\\
$[p_1,q_1,p_2,q_2]$ & & & & & &
\\
\hline
\end{tabular}
\caption{Dynamic transitions table. Looking at
Fig. \ref{fig:dyn_PhDi}, the phases are ordered starting from the
1-F-1RSB one and spiraling counterclockwise around the quadricritical
point. Notice that $p_2$ and $q_2$ in the 2RSB solution and $p_1$ and
$q_1$ in the 1RSB solutions are equivalent notations to $m$ and
$q(1)$, respectively. }
\label{tab:dyn_trans}
\end{table}
\end{center}
\end{widetext}

\section{Conclusion}
In conclusion, developing the analysis of the $s+p$ spin spherical
model with $s,p>2$ and large $p-s$, we find a very rich phase diagram
with various candidate (mean-field) glassy phases, besides the usual
1RSB one.  In a dynamic parallel, a RSB step is connected to a
time-scales bifurcation of processes taking place in the glass former
\cite{Sompolinsky81,CriLeu07}.  In the case of one and two step RSB's,
we have, thus, phases with two or three kinds of processes active on
separated time-scales.  
\\
\indent
In the solid glass formation, the main bifurcation is the one between
the relaxation time of $\alpha$ processes, carrying structural
relaxation and falling out of equilibrium at the glass transition
(i.e., operatively speaking, when the viscosity of the glassformer
reaches the value of $10^{13}$ Poise), and the relaxation times of all
faster (and equilibrated) $\beta$ processes. The difference among the
different existing $\beta$ processes is, usually, neglected in
theoretical modeling, even though their relaxation times can differ
for several orders of magnitude. The 2RSB phase might, then, describe
systems where slow $\beta$ (as, e.g., Johari-Goldstein processes
\cite{JohGol71}) and fast $\beta$ (e.g., collisions) processes are
well separated. This is, indeed, what happens in many glasses where,
at a temperature a few degrees below the glass transition temperature,
Johari-Goldstein-like $\beta$ relaxation occurs on time scales up to
the order of the millisecond. This is much less than the $\alpha$
relaxation time (ca. $10^3$s) but about nine or ten orders of
magnitude larger than the typical times of short-range collision of
the molecules in the viscous liquid phase (the so-called {\em cage
rattling}).
\\
\indent
In the framework of disordered systems theories, moreover, the
existence of a thermodynamically consistent 2RSB phase opens the way
to the study of the complexity contributions at the level of clusters
of states, apart from the standard complexity of states,
allowing for a probe of structures of metastable states in amorphous
systems different from the known ones and including patterns that were
conjectured before but never explicitly computed \cite{Gross85, Montanari03}.
\\ \indent Incidentally, we mention that adding more multibody
interaction terms to the Hamiltonian, Eq. (\ref{f:Ham}), 
\begin{equation}
{\cal H} = \sum_{\alpha=1}^R \sum_{i_1<\ldots <i_{p_\alpha}}J^{(p_\alpha)}_{i_1\ldots i_{p_\alpha}}\sigma_{i_1}
\cdots\sigma_{i_{p_\alpha}}\ ,
\nonumber
\end{equation}
 one is able, in principle, to build a model presenting phases that
are stable within a Parisi Ansatz including any wanted number $R$ of
RSB's.  Indeed, what matters is the shape of the function
$[\Lambda'(q)]^{-1/2}$, the right hand side of
Eq. (\ref{f:solutions}). More precisely, it is the possibility of
finding, for certain, far apart, values of the numbers $p_\alpha$ of
interacting spins and in given regions of the external parameters
space, a $[\Lambda'(q)]^{-1/2}$ whose convexity in the interval $0<q<1$
changes sign a certain number of times.  As we have seen, one change
of convexity allows for the existence of a 2RSB phase. Two changes
would signal the existence of a 3RSB phase, and so forth
\footnote{An example of a model with three multibody interaction terms
is the one whose Langevin dynamics was studied in
Ref. \onlinecite{Caiazzo04}.}.
\\ \indent Eventually, the equivalence of the dynamic equations of
spherical spin-glass models, in the PM phase, with the MC equations of
schematic theories \cite{REV} makes the $s+p$ model also an
interesting instance of an off-equilibrium generalization of the MCT
predictions below the MC transition, where the equivalence breaks
down, since MCT assumes equilibrium (one Gibbs state).  This
would be relevant, above all, to deal with amorphous-to-amorphous
transitions, in which both the interested phases are already frozen
and, thus, equilibrium properties, such as, e.g., the
fluctuation-dissipation theorem, cannot be taken for granted.

\acknowledgments
We acknowledge V. Krakoviack and M. M\"uller for stimulating interactions.


\begin{thebibliography}{99}

\bibitem{Tsiok98} 
  O.B. Tsiok, V.V. Brazhkin, A.G. Lyapin and L.G. Khvostantsev, 
  Phys. Rev. Lett. {\bf 80}, 999 (1998).

\bibitem{Huang04} 
  L. Huang and J. Kieffer, 
  Phys. Rev. B {\bf 69}, 224203 (2004).

\bibitem{Deb01} 
  S.K. Deb, M. Wilding, M.Somayazulu and P.F. McMillian,
  Nature {\bf 414}, 528 (2001).

\bibitem{Mishima85} 
  O. Mishima, L.D. Calvert and E. Walley, 
  Nature {\bf 314}, 74 (1985).

\bibitem{Poole} 
  P.H. Poole, F. Sciortino, U. Essmann and H.E. Stanley,  
  Nature {\bf 360}, 324 (1992).

\bibitem{Chen03} 
  S.-H. Chen, W.-R. Chen and F. Mallamace, 
  Science {\bf 300}, 619 (2003).

\bibitem{Cangialosi05} 
  D. Cangialosi, M. W{\"u}bbenhorst, H. Schut, A. van Veen and S.J. Picken, 
  J. Chem. Phys. {\bf 122}, 064702 (2005). 

\bibitem{Sciortino01} 
  F. Sciortino, 
  Nature materials {\bf 1}, 145 (2002).

\bibitem{DawetAl01} 
  K. Dawson, G. Foffi, M. Fuchs, W. G\"{o}tze,
  F. Sciortino, M. Sperl, P. Tartaglia, Th. Voigtmann, and
  E. Zaccarelli, Phys. Rev. E {\bf 63}, 011401 (2000).

\bibitem{Zac01} 
  E. Zaccarelli, G. Foffi, K.A. Dawson, F. Sciortino and P. Tartaglia, 
  Phys. Rev. E {\bf 63}, 031501 (2001).

\bibitem{Caiazzo04} 
  A. Caiazzo, A. Coniglio, M. Nicodemi, 
  Phys. Rev. Lett.  {\bf 93}, 215701 (2004).  

\bibitem{MPV} M. Mezard, G. Parisi and M.A. Virasoro, 
  {\em Spin glass theory and beyond}, 
  World Scientific (Singapore, 1987).

\bibitem{Gross84} 
  D.J. Gross and M. Mezard,
  Nucl. Phys. B {\bf 240}, 431-452 (1984).

\bibitem{Gardner85} 
  E. Gardner, 
  Nucl. Phys. B {\bf 257}, 747 (1985).

\bibitem{Gross85} 
  D.J. Gross, I. Kanter and H. Sompolinsky, 
  Phys. Rev. Lett. {\bf 55}, 304 (1985). 

\bibitem{SheKir75} D. Sherrington and S. Kirkpatrick, 
Phys. Rev. Lett. {\bf 35}, 1792
(1975).

\bibitem{EdwAnd75} S.F. Edwards and P.W. Anderson, J. Phys. F {\bf 5},
965 (1975).

\bibitem{CriSom92} 
  A. Crisanti and H-.J. Sommers, 
  Z. Phys. B {\bf 87}, 341 (1992).

\bibitem{Montanari03} 
  A. Montanari and F. Ricci-Tersenghi, 
  Eur. Phys. J. B {\bf 33}, 339 (2003).
\bibitem{Zecchina99}
R. Monasson, R. Zecchina, S. Kirkpatrick, B. Selman, L. Troyansky,
Nature {\bf 400}, 133 (1999).
\bibitem{Montanari04} 
  A. Montanari, G. Parisi and F. Ricci-Tersenghi, 
  J. Phys. A {\bf 37}, 2073 (2004).

\bibitem{CriLeu04}
  A. Crisanti and L. Leuzzi,
  Phys. Rev. Lett. {\bf 93}, 217203 (2004).

\bibitem{CriLeu06}
  A. Crisanti and L. Leuzzi,
  Phys. Rev. B {\bf 73}, 014412 (2006).

\bibitem{Nieuwenhuizen95} 
  Th. M. Nieuwenhuizen,
  Phys. Rev. Lett. {\bf 74}, 4289 (1995).

\bibitem{Krakoviack} 
  V. Krakoviack, e-print:  arXiv:0705.3187.

\bibitem{GoeSjo86}
  W. G\"{o}tze and L. Sj\"{o}gren,
  J. Phys. C {\bf 20}, 879 (1986).

\bibitem{GoeSjo89}
  W. G\"{o}tze and L. Sj\"{o}gren,
  J. Phys.: Cond. Matt. {\bf 1}, 4183 (1989); 4203 (1989).

\bibitem{KraAlb97}
  V. Krakoviack, C. Alba-Simionesco and M. Krauzman, 
  J. Chem. Phys. {\bf 107}, 3417 (1997);
\bibitem{KraAlb02}
  V. Krakoviack and C. Alba-Simionesco,
  J. Chem. Phys. {\bf 117}, 2161 (2002).

\bibitem{CiuCri00}
  S. Ciuchi and A. Crisanti,
  Europhys. Lett {\bf 49}, 754 (2000).

\bibitem{CriLeu07} A. Crisanti and L. Leuzzi,
Phys. Rev. B {\bf 75}, 144301 (2007).


\bibitem{CriSom95}
  A. Crisanti and H-.J. Sommers
  J. Phys. I (France) {\bf 5}, 805 (1995).

\bibitem{Monasson95}
  R. Monasson, 
  Phys. Rev. Lett. {\bf 75}, 2847 (1995).  

\bibitem{Sompolinsky81} 
  H. Sompolinsky,
  Phys. Rev. Lett. {\bf 47}, 935 (1981).


\bibitem{REV}
For a review see:
               J. P. Bouchaud, L. F. Cugliandolo, J. Kurchan and M. Mezard
               in {\em Spin glasses and random fields} Ed. by
               A. P. Young (World Scientific, 1997).

\bibitem{RivBir04}  O. Rivoire, G. Biroli, O.C. Martin , M. Mezard,
 Eur. Phys. J. B {\bf 37}, 55 (2004). 

\bibitem{KirTirWol89}
 T.R. Kirkpatrick, D. Thirumalai and P.G.  Wolynes, Phys. Rev. A {\bf 40},
1045 (1989).


\bibitem{CriLeuUnp}
A. Crisanti and L. Leuzzi, unpublished.

\bibitem{Mueller06} M. M\"uller and L. Leuzzi and A. Crisanti, Phys. Rev. B
{\bf 74}, 134431 (2006).

\bibitem{Parisi06} G. Parisi, in {\em Lectures Notes of the Les
Houches Summer School}, A. Bovier, F. Dunlop, A. van Enter, F. Den
Hollander, J. Dalibard eds. (Elsevier, 2006).

\bibitem{JohGol71}
G.P. Johari and M. Goldstein, J. Chem. Phys. {\bf 55}, 4245 (1971).




\bibitem{Crisanti05} A. Crisanti, L. Leuzzi and T. Rizzo, Phys. Rev. B
{\bf 71},  094202  (2005). 


\end{thebibliography}
\end{document}